\documentclass[%
 reprint,
 amsmath,amssymb,
 aps,
]{revtex4-2}

\usepackage{graphicx}% Include figure files
\usepackage{dcolumn}% Align table columns on decimal point
\usepackage{bm}% bold math
\DeclareMathOperator{\sgn}{sgn}

\begin{document}

% \preprint{APS/123-QED}

\title{Alpha buckets in longitudinal phase space: a bifurcation analysis}% Force line breaks with \\

\author{Jernej Frank}
\email{frankj@zedat.fu-berlin.de}
\altaffiliation[Also at ]{Freie Universiteat Berlin, Germany}%Lines break automatically or can be forced with \\
 
\author{Tom Mertens}%

\author{Markus Ries}
 
\affiliation{%
 Helmholtz-Zentrum Berlin fuer Materialien und Energie GmbH (HZB), Berlin, Germany
%  \\
%  This line break forced with \textbackslash\textbackslash
}%

% \collaboration{MUSO Collaboration}%\noaffiliation

%  \homepage{http://www.Second.institution.edu/~Charlie.Author}
% \affiliation{
%  Second institution and/or address\\
%  This line break forced% with \\
% }%
% \affiliation{
%  Third institution, the second for Charlie Author
% \affiliation{%
%  Authors' institution and/or address\\
%  This line break forced with \textbackslash\textbackslash
% }%

% \collaboration{CLEO Collaboration}%\noaffiliation

\date{\today}% It is always \today, today,

\begin{abstract}
At HZB's BESSY II and PTB's Metrology Light Source (MLS) facilities we have the ability to tune the momentum compaction factor $\alpha$ up to second non-linear order. The non-linear dependence $\alpha(\delta)$ brings qualitative changes to the longitudinal phase space and introduces new fix points $\alpha(\delta)=0$ which produce the so-called $\alpha$-buckets. We present with this paper an analysis of this phenomena from the standpoint of bifurcation theory. With this approach we were able to characterize the nature of the fix points and their position in direct dependence on the tunable parameters. Furthermore, we are able to place stringent conditions onto the tunable parameters to either create or destroy $\alpha$-buckets. 
% \begin{description}
% Secondary publications and information retrieval purposes.
% You may use the \texttt{description} environment to structure your abstract;
% use the optional argument of the \verb+\item+ command to give the category of each item. 
% \end{description}
\end{abstract}

\maketitle

%\tableofcontents

\section{\label{sec:intro} Introduction}
% First-level heading:\protect\\ The line
% break was forced \lowercase{via} \textbackslash\textbackslash}
The operation of synchrotron storage rings in the so called quasi-isochronous mode is able to produce shorter bunch lengths leading to short x-ray pulses and coherent sychrotron radiation in the THz radiation regime \cite{chanwattana2016thz, lebasque2012low}. This operation is possible due to the tweaking of the lattice optical parameters to lower the momentum compaction factor $\alpha$. In the low-$\alpha$ mode the linear approximation of the momentum compaction factor is no longer valid and a higher order expansion is necessary. This non-linear behaviour can lead to the so called $\alpha$-buckets \cite{ng1998quasi,murphy2000first}. Placement of sextupole and octupole magnets then enables to tune the first two non-linear orders of $\alpha$ as has been successfully realized at the MLS \cite{feikes2009low} and DLS \cite{martin2011experience}.\\
We neglect path lengthening due to betatron oscillation by assuming a suppressed transverse chromaticity \cite{shoji2005dependence} to obtain a 1-dimensional model and explore the richness of the Hamiltonian.
% to draw a bifurcation diagram. Furthermore, we find Hopf bifurcations occurring to produce further periodic orbits in the longitudinal phase space. We hope to present this theoretical classification of $\alpha$-buckets to show the insight bifurcation theory can provide in manipulating non-linear beam dynamics.\\
In limiting ourselves to the longitudinal phase space we reduce the complexity to a degree where oversight is still possible, but expect that imploring the same principles to the transverse plane will provide a better understanding of the transverse resonant island buckets (TRIBs) cultivated at BESSY II\cite{armborst2020tribs}.\\
In the following we present techniques that are valid in a neighborhood of each fixpoint. In this sense one can treat them as perturbation theory around each fixpoint. Once we move away from the fixpoint and into the neighborhood of another fixpoint the solution around the first fixpoint will become invalid. Global behaviour will then be pieced together by local functions without a smooth (continuous in mathematical language) transition. This of course does not reflect the physics of the Hamiltonian since it is continuous everywhere. Moreover, the analysis performed has to be checked if no "additional physics" has been introduced by the approximation.\\
Another remark we wish to make. We will talk about symmetry breaking and phase transitions of the system as bifurcations. In a simple sense one can understand that as follows: to analyze bifurcations we examine a qualitative change of behaviour in the system by varying the available parameters. For example, if we vary the second order contribution to the momentum compaction factor $\alpha_2$, we can create a Hamiltonian that only has a single fixpoint around the origin of phase space or introduce additional fixpoints next to the origin. By varying the parameters we are modifying the form of the Hamiltonian and either creating or breaking the internal symmetries (conserved quantities also called first integrals like total energy - of course neglecting radiation damping etc.). In this sense one can understand the first integrals as order parameters. By reaching a critical point (or bifurcation point) we break the symmetry of the Hamiltonian, meaning we introduce a phase transition. TRIBs are another great example, one breaks full rotational symmetry of an ellipse to a discrete symmetry when you generate islands.\\
We hope to introduce techniques to the accelerator physics community that have been developed in other areas of physics - predominantly in celestial mechanics. Rather than replace the existing practices we wish to add another vital tool that can be leveraged in non-linear beam dynamics to optimize accelerators. With the theoretical insights it may even be possible to minimally upgrade the existing machines to offer equal radiation source properties sought after in the construction of novel accelerators.\\
The article is organized as follows. Section \ref{sec:phys} gives a physical background to the longitudinal Hamiltonian which we will use to demonstrate the tools of bifurcation theory. Section \ref{sec:bif_theory} gives an introduction into bifurcation theory and presents the tools as an application to the longitudinal Hamiltonian. Section \ref{sec:outlook} provides further directions and gives references to relevant publications that the interested reader may look into to further apply this type of analysis for his respective purposes.

\section{\label{sec:phys}Physical Background}
We begin by introducing the longitudinal dynamics in storage rings. We simply wish to motivate the Hamiltonian and refer the reader to \cite{ries2014nonlinear, wolski2014beam} for a detailed derivation, experimental realization, and the validity of the approximations made. An important note: we will limit the discussion to synhrotron light sources, where the particles move in the ultra-relitivistic limit. This allows us to focus only on the momentum compaction factor $\alpha$. For readers dealing with slower moving particles a relativistic correction is necessary know as the slip factor $\eta$. A calculation of higher order slip factors can be found in \cite{ng2002physics}.

\subsection{Longitudinal Beam Dynamics}
We choose a Frenet-Serret type of moving coordinate system around the reference orbit of the storage ring. The horizontal displacement $x$, vertical displacement $y$, and longitudinal displacement $z$ are measured w.r.t. the reference particle. We introduce energy deviation of a particle as
\begin{equation}
    \delta = \frac{\Delta p}{p_0} = \frac{p-p_0}{p_0} = \frac{1}{\beta^2_0} \frac{\Delta E}{E_0} \approx \frac{\Delta E}{E_0},
\end{equation}
where $p$ is the momentum of our particle, $p_0$ the momentum of the particle moving along the reference orbit, $\beta_0^2 \approx 1$ its velocity, $\Delta E = E - E_0$ the energy deviation of the particle with energy $E$ from the reference particle's energy $E_0$.\\
Particles in storage rings move along with different momenta and thus, on different trajectories. For the closed orbit of a particle we have
\begin{equation}
    L = L_0 (1 + \alpha \delta),
\end{equation}
where $L_0$ is the length of the reference orbit and $\alpha = \alpha(\delta)$ the momentum compaction factor which is momentum dependent. We have the typical expansion up to second order
\begin{align}
    \alpha(\delta) = \alpha_0 + \alpha_1 \delta + \alpha_2 \delta^2,
\end{align}
which we can control at the MLS.\\
Each turn around the ring particles loose energy due to the bending magnets \cite{sands1970physics} and insertion devices \cite{chao1999handbook}. The energy loss is compensated by placement of RF-cavities. Re-accelerating the stored particles gives
\begin{equation}
    \frac{\Delta E}{\text{RF-cavity pass}} = eU(\phi),
\end{equation}
with $e$ the charge of the particle, $U$ the effective accelerating voltage, and $\phi = \phi(t)$ the cavity phase.

\subsection{Hamiltonian}
In the case of the MLS and for the purpose of this paper we will consider a single RF-Cavity where we can model the effective accelerating voltage by
\begin{equation}
    U(\phi) = U_0 \sin \phi,
\end{equation}
with $U_0$ the maximum integrated voltage applied. The general Hamiltonian is constructed as \cite[Eq. 2.39]{ries2014nonlinear}
\begin{equation}
    H(\phi,\delta) = -\beta_0^2 E_0 \int \alpha\ \delta\ d\delta - \frac{eU}{2\pi h} (\cos \phi - \phi \sin \phi_s),
\end{equation}
where $h$ is the harmonic number and $\phi_s$ the synchronous phase.\\
The case of $\alpha$-buckets becomes relevant when approaching the quasi-isochronous operation and the Hamiltonian further simplifies to \cite[Sec. 5]{ries2014nonlinear}
\begin{align}
\label{eq:Hamiltonian}
    H = -A \delta^2 (\frac{\alpha_0}{2} + \frac{\alpha_1}{3} \delta + \frac{\alpha_2}{4} \delta^2) - B \cos \phi,
\end{align}
where $A=\beta_0^2 E_0$ is a fixed property of the lattice, $B(U_0) = \frac{eU_0}{2\pi h}$ can be tuned by the RF-Voltage $U_0$ and $\phi$ is the relative phase of the particle to the RF-cavity phase. The equations of motion then read
\begin{align}
\label{eq:EOM_general}
\begin{split}
    \dot{\phi} &= -A \delta (\alpha_0 + \alpha_1 \delta + \alpha_2 \delta^2)\\
    \dot{\delta} &= -B \sin \phi.
\end{split}
\end{align}

\section{\label{sec:bif_theory}Bifurcation Theory}
The general concern of bifurcation theory is how qualitative properties of the physical system change with respect to the underlying parameter space. For example by tweaking the above parameters in our Hamiltonian (\ref{eq:Hamiltonian}) we can change the nature of fixpoints from stable to unstable and create or annihilate periodic solutions. Fixpoints are points $(\phi_*,\delta_*)$ in phase space such that
\begin{align}
\label{eq:fixpoint}
\begin{split}
     \dot{\phi}(\phi_*,\delta_*) &= 0\\
    \dot{\delta}(\phi_*,\delta_*) &= 0.   
\end{split}
\end{align}
We can then employ perturbation theory around the fixpoints to deduce the nature of the fixpoints, i.e. if they are stable (have periodic solutions around them) or unstable (have hyperbolic behaviour in their neighborhood).\\
We will present four different approaches of determining the nature of fixpoints: Linearization, Morse theory, Hamiltonian potential analysis, and drawing bifurcation diagrams. A detailed treaty on classification and the tools to analyze such behaviour can be found in \cite{verhulst2006nonlinear}.

\subsection{Linearization}
After calculating fixpoints of the non-linear system linearization is the first tool to use to determine if they are stable or unstable. We re-express Eq. \ref{eq:EOM_general} as
\begin{align}
\label{eq:EOM_func}
\begin{split}
    \dot{\phi} &= f(\delta)\\
    \dot{\delta} &= g(\phi).
\end{split}
\end{align}
and extract the theoretical fixpoints by demanding
\begin{align}
\begin{split}
    f(\delta_*) &= 0\\
    g(\phi_*) &= 0,
\end{split}
\end{align}
which yields the six fixpoints 
\begin{align*}
    (\phi_*,\delta_*) = r_* &= (0,0), (\pi,0), (0,\delta_\pm), (\pi,\delta_\pm),\\ \delta_\pm &= \frac{-\alpha_1 \pm \sqrt{\alpha_1^2 - 4 \alpha_0 \alpha_2}}{2\alpha_2}.
\end{align*}
We linearize Eq. \ref{eq:EOM_func} around each fixpoint $r_*$ by
\begin{align}
\begin{split}
    \dot{\phi} &\approx \frac{\partial f}{\partial \delta}(r_*) (\delta - r_*) = -A (\alpha_0 + 2\alpha_1 \delta + 3 \alpha_2 \delta)\big|_{r_*} (\delta - r_*)\\
    \dot{\delta} &\approx \frac{\partial g}{\partial \phi}(r_*) (\phi - r_*) = -B  \cos \phi \big|_{r_*} (\phi - r_*)
\end{split}
\end{align}
and can translate each fixpoint onto the origin
\begin{align}
    \Bar{\phi} &= \phi - r_*\\
    \Bar{\delta} &= \delta - r_*
\end{align}
to obtain
\begin{align}
    \dot{\Bar{\phi}} &\approx \frac{\partial f}{\partial \delta}(r^*) \Bar{\delta} \\
    \dot{\Bar{\delta}} &\approx \frac{\partial g}{\partial \phi}(r^*) \Bar{\phi}.
\end{align}
We will omit the bar over the translated coordinates to not further clutter the notation.\\
If we write it as a matrix we get
\begin{align}
    \begin{pmatrix}
    \dot{\phi}\\
    \dot{\delta}
    \end{pmatrix} = \begin{pmatrix}
    0 & \frac{\partial f}{\partial \delta}(r^*)\\
    \frac{\partial g}{\partial \phi}(r^*) & 0
    \end{pmatrix} \begin{pmatrix}
    \phi\\
    \delta
    \end{pmatrix}
\end{align}
and the eigenvalues follow immediately as
\begin{align}
    \lambda_{1,2} = \pm \sqrt{\frac{\partial f}{\partial \delta}(r^*) \frac{\partial g}{\partial \phi}(r^*)}.
\end{align}
We can have purely imaginary eigenvalues indicating stable periodic solutions or purely real eigenvalues describing hyperbolic behaviour in the neighborhood of the fixpoint. The case of eigenvalues being 0 needs further examination with different tools.\\
For example if we concentrate on the RF-buckets located around the fixpoints $r_1 = (0,0)$ and $r_2 = (\pi,0)$ we obtain
\begin{align}
\label{eq:eig_lin}
\begin{split}
    \lambda_{1,2} (r_1) &= \pm \sqrt{AB\alpha_0}\\
    \lambda_{1,2} (r_2) &= \pm \sqrt{-AB\alpha_0}.
\end{split}
\end{align}
We can see that the fixpoints with $\phi_* = 0$ and $\phi_* = \pi$ will in every case have a sign flip, so if we get a center (periodic) at $\phi_* = 0$ we will get a saddle (hyperbolic) $\phi_* = \pi$. This reduces the analysis for the rest of the fixpoints to only focus on the case $\phi_* = 0$.

\subsection{Morse Theory}
\begin{figure}
    \centering
    \includegraphics[width=\linewidth]{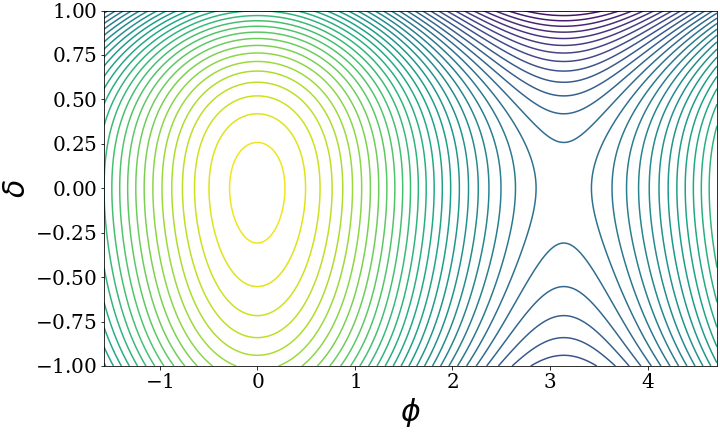}
    \caption{\label{fig:lin_stable_unstable} The energy contours of the Hamiltonian Eq. \ref{eq:Hamiltonian} plotted between $\phi \in [ 0, \pi ]$. We have $AB\alpha_0 < 0$ and can see the stable solution around $(0,0)$ which can be described by a Morse function with index $k=0$ (equation of ellipse). The unstable solution can be seen by either switching the fixpoint $(0,0) \rightarrow (\pi,0)$ as can be deduced from Eq. \ref{eq:eig_lin} or by $AB\alpha_0 \rightarrow - AB\alpha_0$ and can be described by a Morse function with index $k = 1$ (equation of a hyperbola).}
\end{figure}

To give a more general background to the discussion of the previous section on linearization we shortly introduce Morse theory. Consider a more general Hamiltonian function
\begin{align}
    H: \mathbb{R}^{2} \rightarrow \mathbb{R}.
\end{align}
We say $r_*$ is a non-degenerate fixpoint of the Hamiltonian if the determinant of the Hessian matrix
\begin{align}
    \det  \frac{\partial^2 H}{\partial (\phi,\delta)^2} \big |_{r_*} \neq 0.
\end{align}
A smooth function in a neighborhood of a non-degenerate fixpoint $r_*$ is called a Morse function \cite{lang1985differential}. It can be transformed into the form
\begin{align}
    H = H_0 - x_1^2 - ... - x_k^2 + x_{k+1}^2 + ... + x_n^2,
\end{align}
where $k$ is called the index of the Morse function and $H_0$ some constant. Hence, if we are searching for periodic solutions around the fixpoint our analysis reduces to determining the hypersurface in parameter space where the fixpoints are non-degenerate and the Morse index is $k=0$.\\
Let us focus on the fixpoint $r_1 = (0,0)$ again. The Hessian determinant is
\begin{align}
    \det  \frac{\partial^2 H}{\partial (\phi,\delta)^2} \big |_{(0,0)} = -AB\alpha_0
\end{align}
so as long as $AB\alpha_0 \neq 0$ we are dealing with a non-degenerate fixpoint. For $AB\alpha_0 < 0$ we obtain the Morse index $k = 0$ and for $AB\alpha_0 > 0$ the Morse index $k = 1$. We illustrate this behaviour around $\delta_* = 0$ and $\phi_* = 0, \pi$ in Fig. \ref{fig:lin_stable_unstable} by plotting the energy contours of the Hamiltonian Eq. \ref{eq:Hamiltonian}.

\subsection{Hamiltonian Potential Analysis}
\begin{figure*}
    \centering
    \includegraphics[width=.9\linewidth]{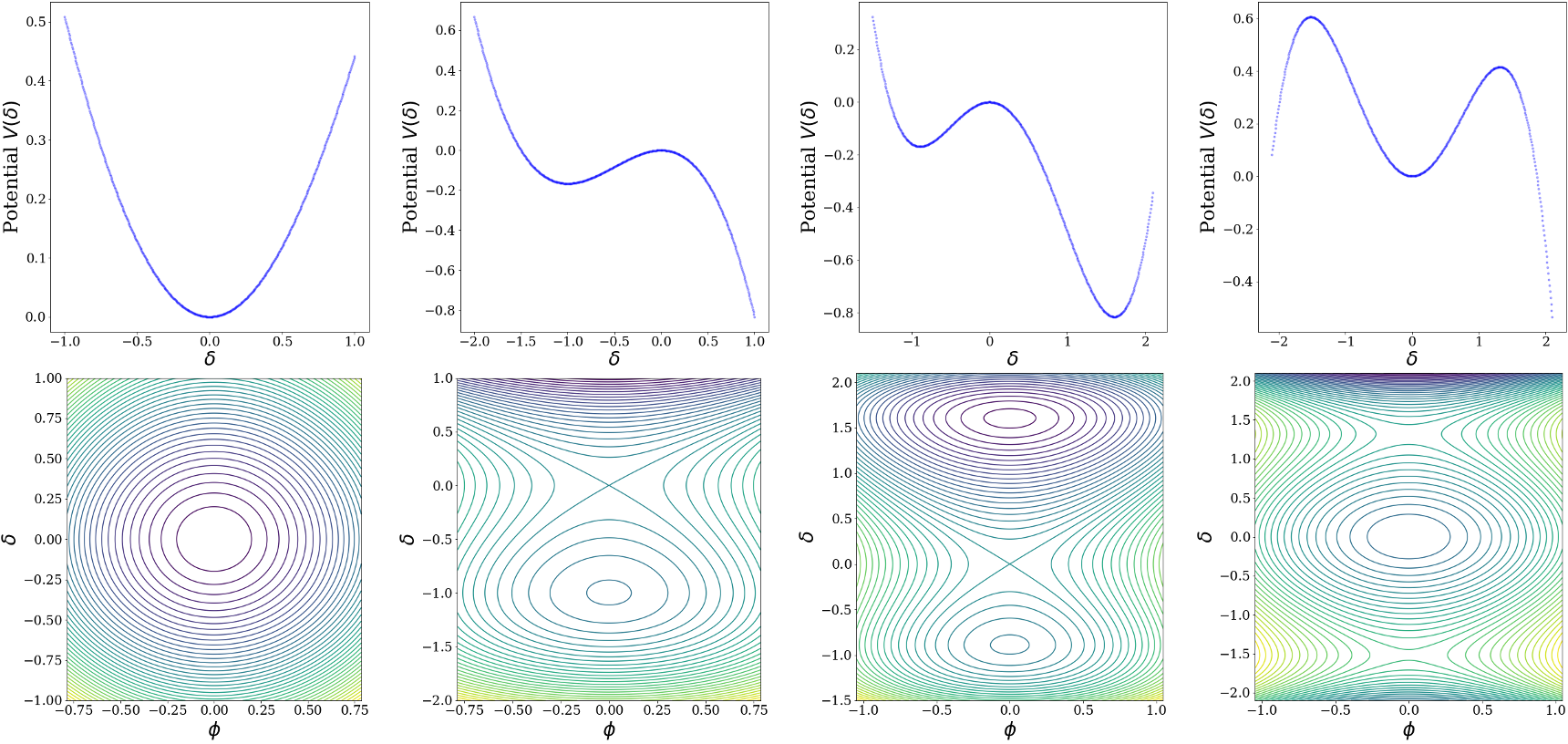}
    \caption{\label{fig:ham_pot_fix} \textbf{Above:} We show the different possible potential functions $V(\delta)$ from Eq. \ref{eq:ham_pot_explicit} for fixed values of the underlying parameters. From left to right: a single potential well which is equivalent to the harmonic oscillator, two extrema - a maximum and a minimum, and three extrema which can be further distinguished by two stable fixpoints or a single stable fixpoint.\textbf{Bellow:} The corresponding energy contours for the Hamiltonian surface corresponding to the Hamiltonian Eq. \ref{eq:hamiltonian_potential}, with the corresponding potential $V(\delta)$ above.}
\end{figure*}

This is a special case of linearization that has a certain elegance to it and may be utilized to gain global insight in a rapid fashion. Consider the special case when the Hamilton function can be written as
\begin{align}
\label{eq:hamiltonian_potential}
H(\phi,\delta) = \frac{\phi^2}{2} + V(\delta)
\end{align}
with potential $V(\delta)$ which can be expanded in a Taylor series in a neighbourhood
of each critical point. An isolated minimum of the potential corresponds with a
stable fixpoint, an isolated maximum corresponds with an unstable
fixpoint \cite[Thm. 8.5]{verhulst2006nonlinear}.\\
The extrema can be easily calculated by
\begin{align}
    \frac{d V(\delta)}{d \delta} = 0
\end{align}
and we can use the second derivative test to obtain
\begin{align}
\begin{split}
    \frac{d^2 V(\delta)}{d\delta^2} &< 0 \quad  \text{maximum},\\
    \frac{d^2 V(\delta)}{d\delta^2} &> 0 \quad  \text{minimum}.
\end{split}
\end{align}
In our Hamiltonian Eq. \ref{eq:Hamiltonian} the potential reads
\begin{align}
\label{eq:ham_pot_explicit}
    V(\delta) = -A (\frac{\alpha_0}{2} \Bar{\delta}^2 + \frac{\alpha_1}{3} \Bar{\delta}^3 + \frac{\alpha_2}{4} \Bar{\delta}^4),
\end{align}
where $\Bar{\delta} = B \delta$ has been re-scaled to obtain the Hamiltonian in the desired form Eq. \ref{eq:hamiltonian_potential}. For the fixpoint $r_* = (0,0)$ we have
\begin{align}
    \frac{d^2 V}{d\delta^2} &< 0 \quad  \text{for } A\alpha_0 B > 0,\\
    \frac{d^2 V}{d\delta^2} &> 0 \quad  \text{for } A\alpha_0 B < 0,
\end{align}
which corresponds exactly to how we determined centers and saddles in linearization.\\
For the remaining two fixpoints (setting $A = B=1$ to reduce the clutter) we have
\begin{align}
    \frac{d^2 V(\delta_+)}{d\delta^2} &= 2\alpha_{0} - \frac{\alpha_{1}^{2}}{2 \alpha_{2}} + \frac{\alpha_{1} \sqrt{\alpha_{1}^{2}- 4 \alpha_{0} \alpha_{2}}}{2 \alpha_{2}}\\
    \frac{d^2 V(\delta_-)}{d\delta^2} &= 2\alpha_{0} - \frac{\alpha_{1}^{2}}{2 \alpha_{2}} - \frac{\alpha_{1} \sqrt{\alpha_{1}^{2}- 4 \alpha_{0} \alpha_{2}}}{2 \alpha_{2}}.
\end{align}
From the general form of the potential we can already deduce that we can have one, two, or three fixpoints in the $\delta$-plane (since we are limiting the discussion around $\phi = 0$) as shown in Fig. \ref{fig:ham_pot_fix}. In the case of three fixpoints, supposing that $(0,0)$ is stable, necessary implies the other two fixpoints are unstable. On the other hand, we can have two centers around $\delta_\pm$ and $(0,0)$ is unstable.

\subsection{Bifurcation Diagrams}
It is now time to turn our attention to the full Hamiltonian Eq. \ref{eq:Hamiltonian} and go beyond linearization to find additional subtleties missed by the above crude approximation. We will focus on the two fixpoints $\phi_* = 0 , \pi$ first. The expansions around the two fixpoints differ only by an overall sign change, which we will absorb into the parameter $B$ and obtain
\begin{align}
    \begin{split}
        \dot{\phi} &= -A(\alpha_0 \delta + \alpha_1 \delta^2 + \alpha_2 \delta^3)\\
        \dot{\delta} &\approx -B (\phi - \frac{1}{6}\phi^3 + ...),
    \end{split}
\end{align}
where $B = \sgn(\phi_*)B$, $\sgn(0) = 1$ and $\sgn(\pi) = -1$. We can now rewrite the above equations as
\begin{align}
    \ddot{\delta} - \zeta (\delta + \lambda \delta^2 + \mu \delta^3) = \epsilon F(\phi, \delta),
\end{align}
with $\zeta = AB\alpha_0,$  $\lambda = \frac{\alpha_1}{\alpha_0},$ $\mu = \frac{\alpha_2}{\alpha_0}$ and treating the higher orders as a perturbation $\epsilon F(\phi, \delta) = -\frac{\zeta}{2} \phi^2 (\delta + \lambda \delta^2 + \mu \delta^3)$.

\subsubsection{Unperturbed Hamiltonian}
\begin{figure}
    \centering
    \includegraphics[width=\linewidth]{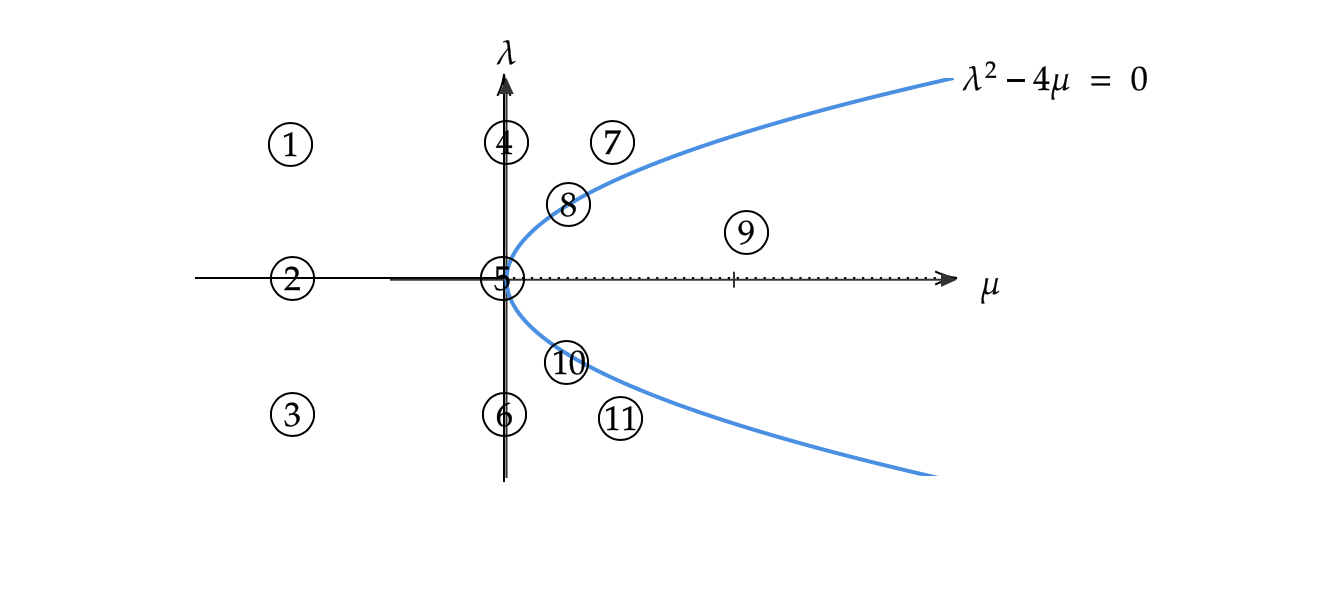}
    \caption{Bifurcation diagram for the remaining two parameters $\lambda, \mu$ with the regions of interest classified by different possible fixpoints Eq. \ref{eq:unperturbed_bifurcation_lines} and marked by numbers 1-11.}
    \label{fig:bifurcation_diagram}
\end{figure}

\begin{figure}
    \centering
    \includegraphics[width=\linewidth]{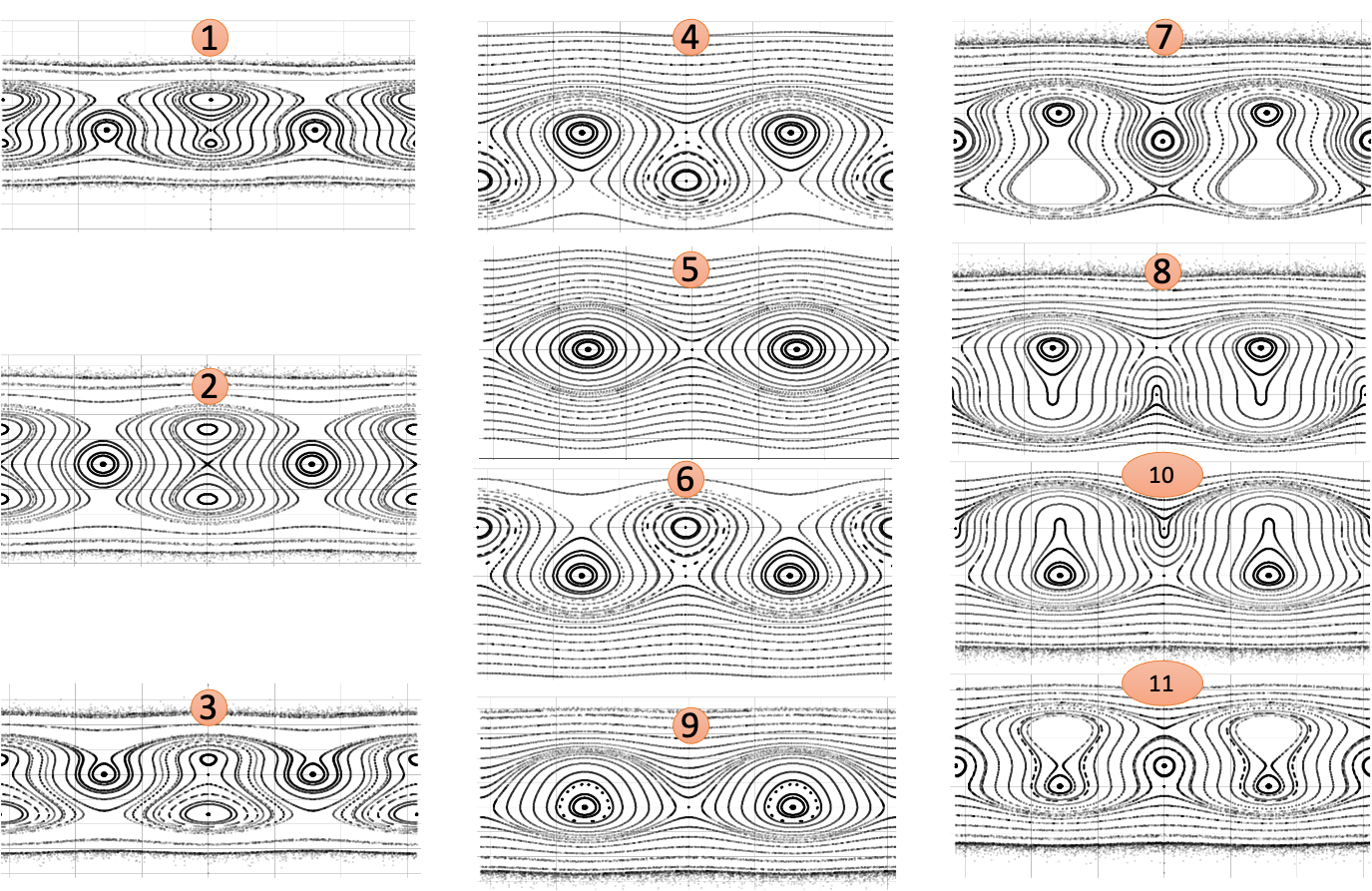}
    \caption{Schematics corresponding to the regions in diagram Fig. \ref{fig:bifurcation_diagram}. We have $\zeta > 0$. The x-axis corresponds to values $\phi \in [-2\pi,2\pi]$ and the y-axis for $\delta \in [-4,4]$ in relative units.}
    \label{fig:bifurcation_track_positiveZ}
\end{figure}

\begin{figure}
    \centering
    \includegraphics[width=\linewidth]{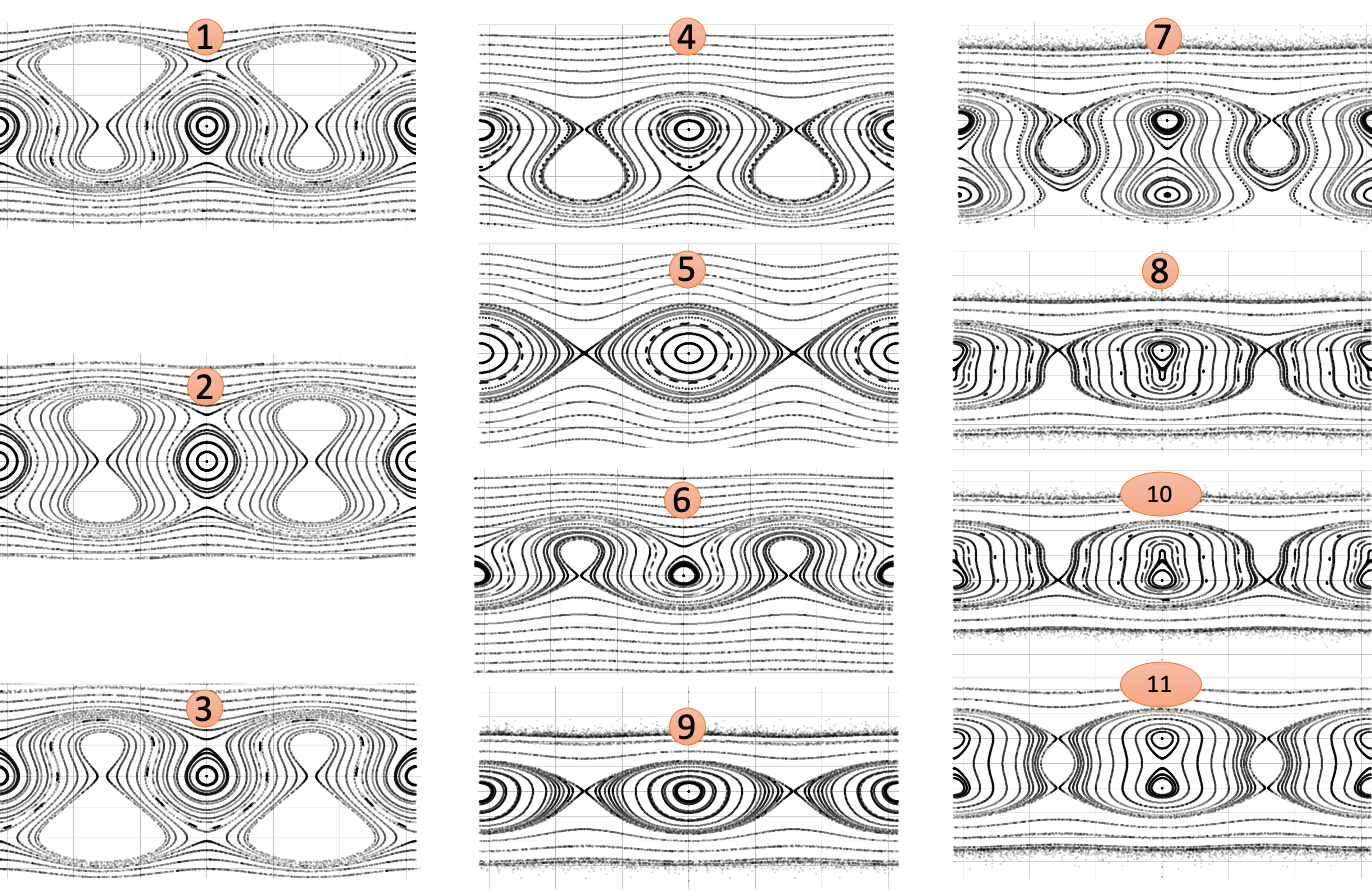}
    \caption{Schematics corresponding to the regions in diagram Fig. \ref{fig:bifurcation_diagram}. We have $\zeta < 0$. The x-axis corresponds to values $\phi \in [-2\pi,2\pi]$ and the y-axis for $\delta \in [-4,4]$ in relative units.}
    \label{fig:bifurcation_track_negativeZ}
\end{figure}

The unperturbed system, $\epsilon = 0$,
\begin{align}
\label{eq:2nd_order_unperturbed}
    \ddot{\delta} - \zeta (\delta + \lambda \delta^2 + \mu \delta^3) = 0.
\end{align}
is the above linearized system. We summarize here the results and show how to categorize them.\\
If $\zeta > 0$ we have the expansion around $\phi_* = 0$ and for $\zeta < 0$ we have the expansion around $\phi_* = \pi$. This is our first global bifurcation example. The system undergoes this bifurcation in 2 ways: either we change $\sgn(\zeta)$ by having a negative momentum compaction factor $\alpha_0$ or reversing the RF-voltage (akin to introducing a $\pi$ phase shift w.r.t. the cavity). As can be easily seen from Eq. \ref{eq:2nd_order_unperturbed}, we either have a harmonic or a hyperbolic solution. Hence, this global bifurcation will result in flipping between stable and unstable fixpoints in our phase plots.\\
The remaining three fixpoints of the system are
\begin{align}
\label{eq:unperturbed_bifurcation_lines}
\begin{split}
    \delta_0 &= 0\\
    \delta_\pm &= \frac{-\lambda \pm \sqrt{\lambda^2 - 4\mu}}{2\mu}, \quad \mu \neq 0\\
    \delta_\pm &= -\frac{1}{\lambda}, \quad \mu =0 .
\end{split}
\end{align}
We have already de-cluttered our equations from 5 parameters down to 3. In principle we could draw a 3D bifurcation diagram, but the $\zeta$-axis would not give us a lot more insight into the behaviour (just mirroring of the other 2 planes along the axis). Therefore, let us draw a 2D bifurcation diagram for the remaining two parameters $\lambda, \mu$ as depicted in Fig. \ref{fig:bifurcation_diagram}.
We have sketched the corresponding particle behaviour in longitudinal phase space for each region and show the global bifurcation $\zeta \rightarrow - \zeta$ by comparing Fig. \ref{fig:bifurcation_track_positiveZ} and Fig. \ref{fig:bifurcation_track_negativeZ}.
\\
We can now take a closer look at regions 1-11 in Figs. \ref{fig:bifurcation_diagram},\ref{fig:bifurcation_track_positiveZ},\ref{fig:bifurcation_track_negativeZ}. For region 5 we have effectively the equation of the harmonic oscillator, depending on the sign of $\zeta$ this will have stable orbits around $0$ and unstable hyperbolic behaviour around $\pi$ and vice-versa. For region 9 the only stable fixpoint is at $(0,0)$ or $(0,\pi)$ (again depending on $\sgn \zeta$) since the remaining fixpoints are imaginary.\\
The region pairs (1,3), (4,6), (7,11), and (8,10) reflect the symmetry of the bifurcation diagram in the $\lambda$-axis. By switching $\lambda\rightarrow -\lambda$ we effectively mirror the behaviour over the fixpoint (0,0) since the position of the stable and unstable fixpoints poses the same symmetry. This symmetry is completely seen in region 2. Since sextupoles are the dominant element to tune $\alpha_1$, if we set them in such a way that $\alpha_1 \approx 0$, we can expand the stable region by pushing the stable fixpoints further apart. Since their positions are then given by $\pm \frac{\sqrt{-\mu}}{\mu}$, one must only be careful not to destroy the envelope region as discussed in the next section on Hopf bifurcations.\\
Regions 4 and 6 in addition demonstrate the behaviour where the contribution of $\alpha_2 \approx 0$, i.e. $\mu=0$.\\
For regions 7 and 11, $\alpha_1^2$ has to dominate over $4\alpha_0 \alpha_2$, meaning the ratio $\frac{\alpha_0 \alpha_2}{\alpha_1^2}$ cannot exceed 25 \%. \\
Regions 8 and 10 demonstrate the limiting behaviour where two fixpoints merge together and become degenerate.
\subsubsection{Hopf Bifurcations}
\begin{figure}
    \centering
    \includegraphics[width=\linewidth]{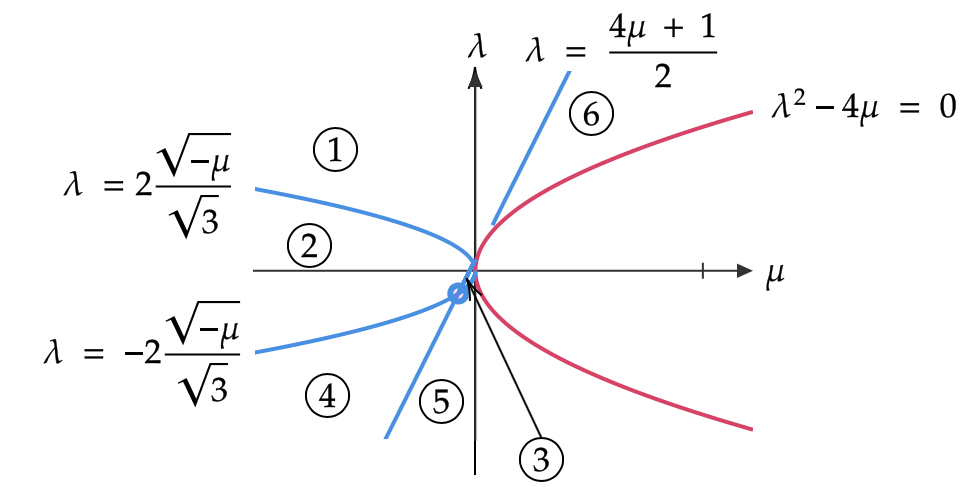}
    \caption{Hopf curves drawn into the bifurcation diagram Fig. \ref{fig:bifurcation_diagram}. We have $\zeta > 0$.}
    \label{fig:hopf_digram}
\end{figure}

\begin{figure}
    \centering
    \includegraphics[width=\linewidth]{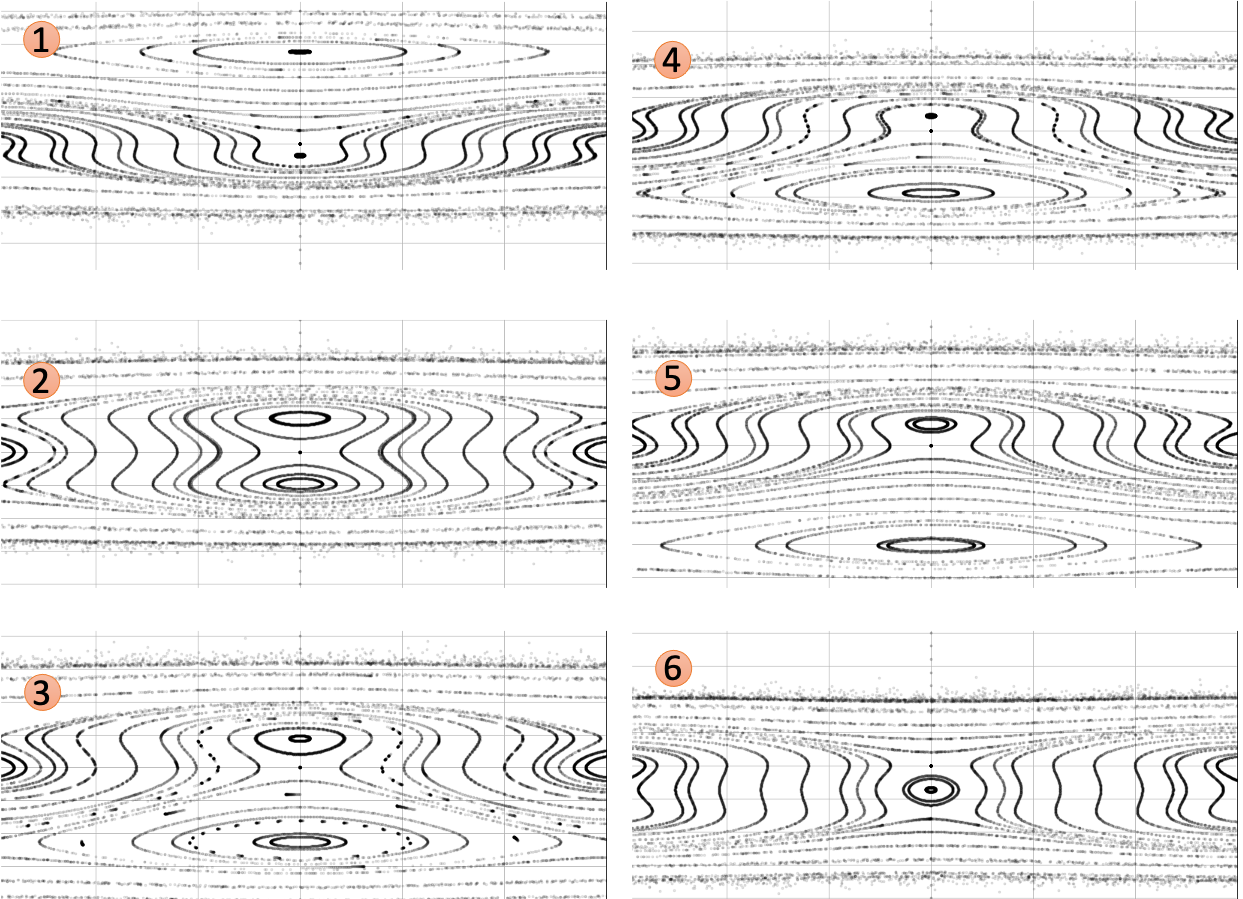}
    \caption{Creation or annihilation of periodic solutions depending on the Hopf region from Fig. \ref{fig:hopf_digram}. We have $\zeta > 0$ and the diagram is magnified only around the region of $\phi = 0$.}
    \label{fig:hopf_tracking}
\end{figure}

\begin{figure}
    \centering
    \includegraphics[width=\linewidth]{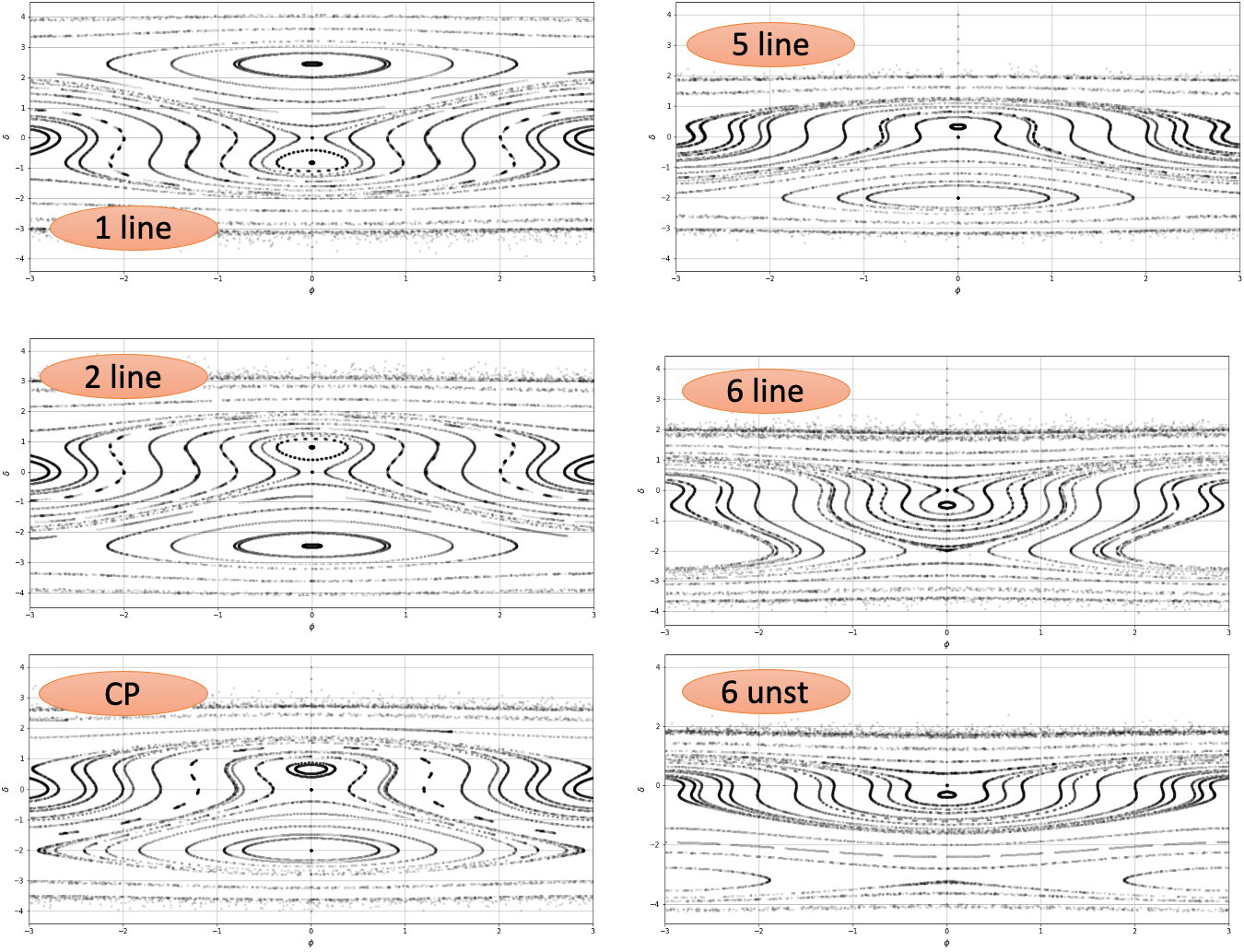}
    \caption{Behaviour on the Hopf lines of Fig. \ref{fig:hopf_digram}. The labels runs as follows: 1 line denotes the Hopf line between region 1 and 2, 2 line between regions 2 and 4, 5 line betwen regions 4 and 5, 6 line next to region 6, CP is the critical point where line 2 and line 5 intersect, and 6 unst the region over the 6 line where periodic solutions from regions are destroyed. We have $\zeta > 0$ and the diagram is magnified only around the region of $\phi = 0$.}
    \label{fig:hopf_marginal}
\end{figure}

Let us now examine the perturbed equation
\begin{align}
\label{eq:unperturbed_delta}
    \ddot{\delta} - \zeta (\delta + \lambda \delta^2 + \mu \delta^3) = \epsilon F(\phi, \delta),
\end{align}
The characteristic polynomial for eigenvalues $l$ at any of the critical points is
\begin{align}
    l^2 + \epsilon \frac{\partial F}{\partial \phi} l + \text{const.} = 0.
\end{align}
We now have to examine the critical points for Hopf bifurcations (where the real part of the eigenvalues becomes zero) to find additional closed orbit solutions. We only need to focus on fixpoints where $\phi_* = 0$ due to the above mentioned global bifurcation $\zeta \rightarrow - \zeta$. \\
For $\alpha$-buckets the above characteristic equation does not help us since
\begin{align}
    \frac{\partial F}{\partial \phi} = 0
\end{align}
for any values of $\lambda,\mu$ and we cannot deduce any information from it. We have to resort to higher order expansions beyond linearization of the second order differential equation (\ref{eq:unperturbed_delta}). The first non-trivial derivative yields more information. The fixpoint $\delta_* = 0$ is trivial. If we insert the fixpoints $\delta_\pm$ we get
\begin{align}
\begin{split}
    \frac{\partial^3 F(\phi,\delta)}{\partial \phi^2 \partial \delta} \big |_{(0,\delta_\pm)} &= -4\mu + 3\lambda^2 - 2\lambda \\
    &\pm \sqrt{\lambda^2 - 4\mu} \mp 3\lambda\sqrt{\lambda^2-4\mu} \stackrel{!}{=} 0.
\end{split}
\end{align}
With this we are able to amend our previous bifurcation diagram Fig. \ref{fig:bifurcation_diagram} with the lines for Hopf bifurcations as depicted in Fig. \ref{fig:hopf_digram}.
We have again sketched the corresponding particle behaviour for the regions of interest as can be seen in Fig. \ref{fig:hopf_tracking}.
We see that by crossing from region $1 \rightarrow 2$ our system obtains additional periodic solutions around the stable fixpoints $\delta_\pm$ creating an envelope. The envelope connects the two buckets and allows for particle redistribution.
% Such an occurrence may be used as a lifetime feedback "re-injecting" a particle from the lower into the upper bucket (for comp. \cite[Fig. 5.14]{ries2014nonlinear}) to exactly compensate the lifetime
% losses.\\
The marginal cases on the Hopf lines of Fig. \ref{fig:hopf_digram} show that by crossing the line of region 6 we destroy periodic solutions as depicted in Fig. \ref{fig:hopf_marginal}.
\section{\label{sec:outlook}Conclusion and Outlook}
With this contribution we hope to have presented bifurcation theory as a powerful tool to be employed in the accelerator physics community to control non-linear beam dynamics. The fundamental shift from phase space to parameter space allows us to decide which particular non-linear effect we wish to employ in the accelerator and then find the corresponding hypersurface in parameter space where this is possible. Furthermore, we are also able to place stringent conditions between parameters as algebraic constraints which allows us to maximize or minimize effects.\\
We have introduced on the example of $\alpha$-buckets linearization as a tool to examine the fixpoints. With it we can deduce their position and if they are centers or saddles as long as we are dealing with non-degenerate cases (i.e. Hessian determinant is unequal to 0). Furthermore, we were able to see how the individual fixpoints are linked together in a kind of dual behaviour (e.g. the fixpoints around $\phi = 0$ and $\phi = \pi$). By using this crude tool, we can already obtain a sense of how to change the parameters to produce the desired effect as for example first demanding the fixpoints to be at a certain position and then using the remaining freedom to tune the parameters in such a way, that the placed fixpoints are either centers or saddles.\\
We note that the Hamiltonian potential analysis can only be performed locally since the Taylor expansion is local around a certain fixpoint. However, if we do examine the nature of the fixpoints in this manner, we start to manipulate  $\alpha$-buckets by putting algebraic constraints on the the parameter space.\\ 
Naturally, the higher order neglected terms also contribute to the physics so we cannot neglect them completely. We have shown that creation of additional periodic solutions missed by leading order approximations is determined by the condition of Hopf bifurcations to occur. As an interesting extension we propose to find homoclinic bifurcations by constructing so called Poincaré mappings in the spirit of \cite[chap. 4]{doelman1994bifurcations}.\\
The interested reader may want to look into \cite{reithmeier2006periodic,verhulst2006nonlinear} for a more applied approach to presenting the subject, \cite{sanders2007averaging, arnold2013dynamical, arnold2012geometrical} for a more mathematical introduction, and \cite{sadovskii1996bifurcation} for an application using normal forms and resonance theory.

% \begin{acknowledgments}
% We are grateful to the colleagues of the HZB accelerator complex for their critical discussions during the preparation of this paper.
% \end{acknowledgments}

\appendix

\nocite{*}

\bibliography{bif_ab}% Produces the bibliography via BibTeX.

\providecommand{\noopsort}[1]{}\providecommand{\singleletter}[1]{#1}%
\begin{thebibliography}{21}%
\makeatletter
\providecommand \@ifxundefined [1]{%
 \@ifx{#1\undefined}
}%
\providecommand \@ifnum [1]{%
 \ifnum #1\expandafter \@firstoftwo
 \else \expandafter \@secondoftwo
 \fi
}%
\providecommand \@ifx [1]{%
 \ifx #1\expandafter \@firstoftwo
 \else \expandafter \@secondoftwo
 \fi
}%
\providecommand \natexlab [1]{#1}%
\providecommand \enquote  [1]{``#1''}%
\providecommand \bibnamefont  [1]{#1}%
\providecommand \bibfnamefont [1]{#1}%
\providecommand \citenamefont [1]{#1}%
\providecommand \href@noop [0]{\@secondoftwo}%
\providecommand \href [0]{\begingroup \@sanitize@url \@href}%
\providecommand \@href[1]{\@@startlink{#1}\@@href}%
\providecommand \@@href[1]{\endgroup#1\@@endlink}%
\providecommand \@sanitize@url [0]{\catcode `\\12\catcode `\$12\catcode
  `\&12\catcode `\#12\catcode `\^12\catcode `\_12\catcode `\%12\relax}%
\providecommand \@@startlink[1]{}%
\providecommand \@@endlink[0]{}%
\providecommand \url  [0]{\begingroup\@sanitize@url \@url }%
\providecommand \@url [1]{\endgroup\@href {#1}{\urlprefix }}%
\providecommand \urlprefix  [0]{URL }%
\providecommand \Eprint [0]{\href }%
\providecommand \doibase [0]{https://doi.org/}%
\providecommand \selectlanguage [0]{\@gobble}%
\providecommand \bibinfo  [0]{\@secondoftwo}%
\providecommand \bibfield  [0]{\@secondoftwo}%
\providecommand \translation [1]{[#1]}%
\providecommand \BibitemOpen [0]{}%
\providecommand \bibitemStop [0]{}%
\providecommand \bibitemNoStop [0]{.\EOS\space}%
\providecommand \EOS [0]{\spacefactor3000\relax}%
\providecommand \BibitemShut  [1]{\csname bibitem#1\endcsname}%
\let\auto@bib@innerbib\@empty
%</preamble>
\bibitem [{\citenamefont {Chanwattana}\ \emph {et~al.}(2016)\citenamefont
  {Chanwattana}, \citenamefont {Atay}, \citenamefont {Bartolini}, \citenamefont
  {Cinque}, \citenamefont {Frogley}, \citenamefont {Koukovini-Platia},\ and\
  \citenamefont {Martin}}]{chanwattana2016thz}%
  \BibitemOpen
  \bibfield  {author} {\bibinfo {author} {\bibfnamefont {T.}~\bibnamefont
  {Chanwattana}}, \bibinfo {author} {\bibfnamefont {M.}~\bibnamefont {Atay}},
  \bibinfo {author} {\bibfnamefont {R.}~\bibnamefont {Bartolini}}, \bibinfo
  {author} {\bibfnamefont {G.}~\bibnamefont {Cinque}}, \bibinfo {author}
  {\bibfnamefont {M.}~\bibnamefont {Frogley}}, \bibinfo {author} {\bibfnamefont
  {E.}~\bibnamefont {Koukovini-Platia}},\ and\ \bibinfo {author} {\bibfnamefont
  {I.}~\bibnamefont {Martin}},\ }\bibfield  {title} {\bibinfo {title} {Thz
  coherent synchrotron radiation from ultra-low alpha operating mode at diamond
  light source},\ }\href@noop {} {\bibfield  {journal} {\bibinfo  {journal}
  {Proc. IPAC’16}\ ,\ \bibinfo {pages} {1682}} (\bibinfo {year}
  {2016})}\BibitemShut {NoStop}%
\bibitem [{\citenamefont {Lebasque}(2012)}]{lebasque2012low}%
  \BibitemOpen
  \bibfield  {author} {\bibinfo {author} {\bibfnamefont {J.-P.~P.}\
  \bibnamefont {Lebasque}},\ }\href@noop {} {\bibinfo {title} {Low-alpha
  operation for the soleil storage ring}} (\bibinfo {year} {2012})\BibitemShut
  {NoStop}%
\bibitem [{\citenamefont {Ng}(1998)}]{ng1998quasi}%
  \BibitemOpen
  \bibfield  {author} {\bibinfo {author} {\bibfnamefont {K.-Y.}\ \bibnamefont
  {Ng}},\ }\bibfield  {title} {\bibinfo {title} {Quasi-isochronous buckets in
  storage rings},\ }\href@noop {} {\bibfield  {journal} {\bibinfo  {journal}
  {Nuclear Instruments and Methods in Physics Research Section A: Accelerators,
  Spectrometers, Detectors and Associated Equipment}\ }\textbf {\bibinfo
  {volume} {404}},\ \bibinfo {pages} {199} (\bibinfo {year}
  {1998})}\BibitemShut {NoStop}%
\bibitem [{\citenamefont {Murphy}\ and\ \citenamefont
  {Kramer}(2000)}]{murphy2000first}%
  \BibitemOpen
  \bibfield  {author} {\bibinfo {author} {\bibfnamefont {J.}~\bibnamefont
  {Murphy}}\ and\ \bibinfo {author} {\bibfnamefont {S.}~\bibnamefont
  {Kramer}},\ }\bibfield  {title} {\bibinfo {title} {First observation of
  simultaneous alpha buckets in a quasi-isochronous storage ring},\ }\href@noop
  {} {\bibfield  {journal} {\bibinfo  {journal} {Physical review letters}\
  }\textbf {\bibinfo {volume} {84}},\ \bibinfo {pages} {5516} (\bibinfo {year}
  {2000})}\BibitemShut {NoStop}%
\bibitem [{\citenamefont {Feikes}\ \emph {et~al.}(2009)\citenamefont {Feikes},
  \citenamefont {von Hartrott}, \citenamefont {W{\"u}stefeld}, \citenamefont
  {Hoehl}, \citenamefont {Klein}, \citenamefont {Muller},\ and\ \citenamefont
  {Ulm}}]{feikes2009low}%
  \BibitemOpen
  \bibfield  {author} {\bibinfo {author} {\bibfnamefont {J.}~\bibnamefont
  {Feikes}}, \bibinfo {author} {\bibfnamefont {M.}~\bibnamefont {von
  Hartrott}}, \bibinfo {author} {\bibfnamefont {G.}~\bibnamefont
  {W{\"u}stefeld}}, \bibinfo {author} {\bibfnamefont {A.}~\bibnamefont
  {Hoehl}}, \bibinfo {author} {\bibfnamefont {R.}~\bibnamefont {Klein}},
  \bibinfo {author} {\bibfnamefont {R.}~\bibnamefont {Muller}},\ and\ \bibinfo
  {author} {\bibfnamefont {G.}~\bibnamefont {Ulm}},\ }\bibfield  {title}
  {\bibinfo {title} {Low alpha operation of the mls electron storage ring},\
  }in\ \href@noop {} {\emph {\bibinfo {booktitle} {Proc. of the 2009 Particle
  Accelerator Conference}}}\ (\bibinfo {year} {2009})\BibitemShut {NoStop}%
\bibitem [{\citenamefont {Martin}\ \emph {et~al.}(2011)\citenamefont {Martin},
  \citenamefont {Rehm}, \citenamefont {Thomas},\ and\ \citenamefont
  {Bartolini}}]{martin2011experience}%
  \BibitemOpen
  \bibfield  {author} {\bibinfo {author} {\bibfnamefont {I.}~\bibnamefont
  {Martin}}, \bibinfo {author} {\bibfnamefont {G.}~\bibnamefont {Rehm}},
  \bibinfo {author} {\bibfnamefont {C.}~\bibnamefont {Thomas}},\ and\ \bibinfo
  {author} {\bibfnamefont {R.}~\bibnamefont {Bartolini}},\ }\bibfield  {title}
  {\bibinfo {title} {Experience with low-alpha lattices at the diamond light
  source},\ }\href@noop {} {\bibfield  {journal} {\bibinfo  {journal} {Physical
  Review Special Topics-Accelerators and Beams}\ }\textbf {\bibinfo {volume}
  {14}},\ \bibinfo {pages} {040705} (\bibinfo {year} {2011})}\BibitemShut
  {NoStop}%
\bibitem [{\citenamefont {Shoji}(2005)}]{shoji2005dependence}%
  \BibitemOpen
  \bibfield  {author} {\bibinfo {author} {\bibfnamefont {Y.}~\bibnamefont
  {Shoji}},\ }\bibfield  {title} {\bibinfo {title} {Dependence of average path
  length betatron motion in a storage ring},\ }\href@noop {} {\bibfield
  {journal} {\bibinfo  {journal} {Physical Review Special Topics-Accelerators
  and Beams}\ }\textbf {\bibinfo {volume} {8}},\ \bibinfo {pages} {094001}
  (\bibinfo {year} {2005})}\BibitemShut {NoStop}%
\bibitem [{\citenamefont {Armborst}(2020)}]{armborst2020tribs}%
  \BibitemOpen
  \bibfield  {author} {\bibinfo {author} {\bibfnamefont {F.}~\bibnamefont
  {Armborst}},\ }\emph {\bibinfo {title} {Transverse Resonance Island Buckets
  at BESSY II - A new Bunch Separation Scheme -}},\ \href@noop {} {Ph.D.
  thesis},\ \bibinfo  {school} {Humboldt-Universit{\"a}t zu Berlin,
  Mathematisch-Naturwissenschaftliche Fakult{\"a}t I} (\bibinfo {year}
  {2020})\BibitemShut {NoStop}%
\bibitem [{\citenamefont {Ries}(2014)}]{ries2014nonlinear}%
  \BibitemOpen
  \bibfield  {author} {\bibinfo {author} {\bibfnamefont {M.}~\bibnamefont
  {Ries}},\ }\emph {\bibinfo {title} {Nonlinear momentum compaction and
  coherent synchrotron radiation at the metrology light source}},\ \href@noop
  {} {Ph.D. thesis},\ \bibinfo  {school} {Humboldt-Universit{\"a}t zu Berlin,
  Mathematisch-Naturwissenschaftliche Fakult{\"a}t I} (\bibinfo {year}
  {2014})\BibitemShut {NoStop}%
\bibitem [{\citenamefont {Wolski}(2014)}]{wolski2014beam}%
  \BibitemOpen
  \bibfield  {author} {\bibinfo {author} {\bibfnamefont {A.}~\bibnamefont
  {Wolski}},\ }\href@noop {} {\emph {\bibinfo {title} {Beam dynamics in high
  energy particle accelerators}}}\ (\bibinfo  {publisher} {World Scientific},\
  \bibinfo {year} {2014})\BibitemShut {NoStop}%
\bibitem [{\citenamefont {Ng}(2002)}]{ng2002physics}%
  \BibitemOpen
  \bibfield  {author} {\bibinfo {author} {\bibfnamefont {K.-Y.}\ \bibnamefont
  {Ng}},\ }\href@noop {} {\emph {\bibinfo {title} {Physics of intensity
  dependent beam instabilities}}},\ \bibinfo {type} {Tech. Rep.}\ (\bibinfo
  {year} {2002})\BibitemShut {NoStop}%
\bibitem [{\citenamefont {Sands}(1970)}]{sands1970physics}%
  \BibitemOpen
  \bibfield  {author} {\bibinfo {author} {\bibfnamefont {M.}~\bibnamefont
  {Sands}},\ }\href@noop {} {\emph {\bibinfo {title} {Physics Of Electron
  Storage Rings: An Introduction.}}},\ \bibinfo {type} {Tech. Rep.}\ (\bibinfo
  {institution} {Stanford Linear Accelerator Center, Calif.},\ \bibinfo {year}
  {1970})\BibitemShut {NoStop}%
\bibitem [{\citenamefont {Chao}\ and\ \citenamefont
  {Tigner}(1999)}]{chao1999handbook}%
  \BibitemOpen
  \bibfield  {author} {\bibinfo {author} {\bibfnamefont {A.}~\bibnamefont
  {Chao}}\ and\ \bibinfo {author} {\bibfnamefont {M.}~\bibnamefont {Tigner}},\
  }\href@noop {} {\bibinfo {title} {Handbook of accelerator and engineering}}
  (\bibinfo {year} {1999})\BibitemShut {NoStop}%
\bibitem [{\citenamefont {Verhulst}(2006)}]{verhulst2006nonlinear}%
  \BibitemOpen
  \bibfield  {author} {\bibinfo {author} {\bibfnamefont {F.}~\bibnamefont
  {Verhulst}},\ }\href@noop {} {\emph {\bibinfo {title} {Nonlinear differential
  equations and dynamical systems}}}\ (\bibinfo  {publisher} {Springer Science
  \& Business Media},\ \bibinfo {year} {2006})\BibitemShut {NoStop}%
\bibitem [{\citenamefont {Lang}(1985)}]{lang1985differential}%
  \BibitemOpen
  \bibfield  {author} {\bibinfo {author} {\bibfnamefont {S.}~\bibnamefont
  {Lang}},\ }\href@noop {} {\emph {\bibinfo {title} {Differential manifolds}}}\
  (\bibinfo  {publisher} {Springer},\ \bibinfo {year} {1985})\BibitemShut
  {NoStop}%
\bibitem [{\citenamefont {Doelman}\ and\ \citenamefont
  {Verhulst}(1994)}]{doelman1994bifurcations}%
  \BibitemOpen
  \bibfield  {author} {\bibinfo {author} {\bibfnamefont {A.}~\bibnamefont
  {Doelman}}\ and\ \bibinfo {author} {\bibfnamefont {F.}~\bibnamefont
  {Verhulst}},\ }\bibfield  {title} {\bibinfo {title} {Bifurcations of strongly
  non-linear self-excited oscillations},\ }\href@noop {} {\bibfield  {journal}
  {\bibinfo  {journal} {Mathematical Methods in the applied sciences}\ }\textbf
  {\bibinfo {volume} {17}},\ \bibinfo {pages} {189} (\bibinfo {year}
  {1994})}\BibitemShut {NoStop}%
\bibitem [{\citenamefont {Reithmeier}(2006)}]{reithmeier2006periodic}%
  \BibitemOpen
  \bibfield  {author} {\bibinfo {author} {\bibfnamefont {E.}~\bibnamefont
  {Reithmeier}},\ }\href@noop {} {\emph {\bibinfo {title} {Periodic solutions
  of nonlinear dynamical systems: numerical computation, stability, bifurcation
  and transition to chaos}}}\ (\bibinfo  {publisher} {Springer},\ \bibinfo
  {year} {2006})\BibitemShut {NoStop}%
\bibitem [{\citenamefont {Sanders}\ \emph {et~al.}(2007)\citenamefont
  {Sanders}, \citenamefont {Verhulst},\ and\ \citenamefont
  {Murdock}}]{sanders2007averaging}%
  \BibitemOpen
  \bibfield  {author} {\bibinfo {author} {\bibfnamefont {J.~A.}\ \bibnamefont
  {Sanders}}, \bibinfo {author} {\bibfnamefont {F.}~\bibnamefont {Verhulst}},\
  and\ \bibinfo {author} {\bibfnamefont {J.}~\bibnamefont {Murdock}},\
  }\href@noop {} {\emph {\bibinfo {title} {Averaging methods in nonlinear
  dynamical systems}}},\ Vol.~\bibinfo {volume} {59}\ (\bibinfo  {publisher}
  {Springer},\ \bibinfo {year} {2007})\BibitemShut {NoStop}%
\bibitem [{\citenamefont {Arnold}\ \emph {et~al.}(2013)\citenamefont {Arnold},
  \citenamefont {Afrajmovich}, \citenamefont {Il'yashenko},\ and\ \citenamefont
  {Shil'nikov}}]{arnold2013dynamical}%
  \BibitemOpen
  \bibfield  {author} {\bibinfo {author} {\bibfnamefont {V.~I.}\ \bibnamefont
  {Arnold}}, \bibinfo {author} {\bibfnamefont {V.}~\bibnamefont {Afrajmovich}},
  \bibinfo {author} {\bibfnamefont {Y.~S.}\ \bibnamefont {Il'yashenko}},\ and\
  \bibinfo {author} {\bibfnamefont {L.}~\bibnamefont {Shil'nikov}},\
  }\href@noop {} {\emph {\bibinfo {title} {Dynamical systems V: bifurcation
  theory and catastrophe theory}}},\ Vol.~\bibinfo {volume} {5}\ (\bibinfo
  {publisher} {Springer Science \& Business Media},\ \bibinfo {year}
  {2013})\BibitemShut {NoStop}%
\bibitem [{\citenamefont {Arnold}(2012)}]{arnold2012geometrical}%
  \BibitemOpen
  \bibfield  {author} {\bibinfo {author} {\bibfnamefont {V.~I.}\ \bibnamefont
  {Arnold}},\ }\href@noop {} {\emph {\bibinfo {title} {Geometrical methods in
  the theory of ordinary differential equations}}},\ Vol.\ \bibinfo {volume}
  {250}\ (\bibinfo  {publisher} {Springer Science \& Business Media},\ \bibinfo
  {year} {2012})\BibitemShut {NoStop}%
\bibitem [{\citenamefont {Sadovskii}\ and\ \citenamefont
  {Delos}(1996)}]{sadovskii1996bifurcation}%
  \BibitemOpen
  \bibfield  {author} {\bibinfo {author} {\bibfnamefont {D.}~\bibnamefont
  {Sadovskii}}\ and\ \bibinfo {author} {\bibfnamefont {J.}~\bibnamefont
  {Delos}},\ }\bibfield  {title} {\bibinfo {title} {Bifurcation of the periodic
  orbits of hamiltonian systems: An analysis using normal form theory},\
  }\href@noop {} {\bibfield  {journal} {\bibinfo  {journal} {Physical Review
  E}\ }\textbf {\bibinfo {volume} {54}},\ \bibinfo {pages} {2033} (\bibinfo
  {year} {1996})}\BibitemShut {NoStop}%
\end{thebibliography}%

\end{document}